\PassOptionsToPackage{table}{xcolor}
\documentclass[acmsmall,nonacm]{acmart}

\usepackage{orcidlink} 

\usepackage{longtable, booktabs, multirow, array, ragged2e}
\usepackage[table]{xcolor}
\usepackage{tabularx}
\usepackage{makecell}
\usepackage{rotating}
\usepackage{float}
\usepackage{csquotes}

\usepackage{tikz}

\usepackage{enumitem}
\usepackage[normalem]{ulem}

\usepackage{comment}

\definecolor{headercolor}{RGB}{60,60,60}
\definecolor{lightgray}{RGB}{245,245,245}

\AtBeginDocument{%
  }

\setcopyright{cc}
\setcctype{by}
\acmJournal{PACMHCI}
\acmYear{2025} \acmVolume{9} \acmNumber{7}
\acmArticle{CSCW417} \acmMonth{11}
\acmDOI{10.1145/3757598}

\begin{document}

\title{Families' Vision of Generative AI Agents for Household Safety Against Digital and Physical Threats}

\author{Zikai Wen}
\email{zkwen@uw.edu}
\orcid{0000-0001-9163-7450}
\affiliation{
    \institution{School of Engineering and Technology, University of Washington}
    \city{Tacoma}
    \state{Washington}
    \country{USA}
}

\author{Lanjing Liu}
\email{lliu153@jhu.edu}
\orcid{0000-0003-2723-722X}
\affiliation{
    \institution{Department of Computer Science, Johns Hopkins University}
    \city{Baltimore}
    \state{Maryland}
    \country{USA}
}

\author{Yaxing Yao}
\email{yaxing@jhu.edu}
\orcid{0009-0008-7900-9265}
\affiliation{
    \institution{Department of Computer Science, Johns Hopkins University}
    \city{Baltimore}
    \state{Maryland}
    \country{USA}
}

\begin{abstract}
  As families face increasingly complex safety challenges in digital and physical environments, generative AI (GenAI) presents new opportunities to support household safety through multiple specialized AI agents. Through a two-phase qualitative study consisting of individual interviews and collaborative sessions with 13 parent-child dyads, we explored families' conceptualizations of GenAI and their envisioned use of AI agents in daily family life. Our findings reveal that families preferred to distribute safety-related support across multiple AI agents, each embodying a familiar caregiving role: a household manager coordinating routine tasks and mitigating risks such as digital fraud and home accidents; a private tutor providing personalized educational support, including safety education; and a family therapist offering emotional support to address sensitive safety issues such as cyberbullying and digital harassment. Families emphasized the need for agent-specific privacy boundaries, recognized generational differences in trust toward AI agents, and stressed the importance of maintaining open family communication alongside the assistance of AI agents. Based on these findings, we propose a multi-agent system design featuring four privacy-preserving principles: memory segregation, conversational consent, selective data sharing, and progressive memory management to help balance safety, privacy, and autonomy within family contexts.
\end{abstract}



\received{October 2024}
\received[revised]{April 2025}
\received[accepted]{August 2025}

\maketitle

\section{Introduction}
Families face increasingly complex safety concerns across digital and physical environments~\cite{gottfried_teens_2023}. Digital safety threats include online harassment, cyberbullying, sexual violence, financial fraud, digital addiction, misinformation, and unsafe and illegal behaviors~\cite{freed_understanding_2023}. Physical safety involves concerns such as traffic safety~\cite{niebuhr_pedestrian_2016}, home security~\cite{r_design_2023}, and health emergencies~\cite{papachristou_is_2022}. These safety issues often evolve rapidly and remain hidden from parental supervision~\cite{wang_protection_2021, corcoran_associations_2022}, particularly in digital spaces where children may be reluctant to share their worries with parents~\cite{wisniewski_parents_2017, hernandez_parents_2022}. This gap between parents and children highlights challenges in current safety management methods.

Current family safety management approaches, such as traditional monitoring tools and parental controls, can create power imbalances, leading to children's resistance and evasive behaviors~\cite{wisniewski_parental_2017,ghosh_matter_2018}. Children may hide critical information from parents due to fear of judgment or consequences~\cite{petronio_communication_2010, hernandez_parents_2022}, while parents often struggle with effectively communicating safety concerns or providing timely guidance when needed~\cite{modecki_what_2022,park_its_2024}.

While families do not yet perceive generative AI (GenAI) as an explicit solution for managing household safety, there are conceptual reasons why GenAI models could become an important part of the family safety management toolkit. Recent research has shown that GenAI technologies, with their natural language interactions and adaptive capabilities beyond traditional digital solutions, can provide children with more engaging educational support~\cite{zhang_mathemyths_2024, newman2024want} and companionship~\cite{constantinescu_children-robot_2022, abbasi2025longitudinal}. For instance, Abbasi et al.~\cite{abbasi2025longitudinal} found that some children, particularly girls, rated the AI-powered social robot highly as a confidante, suggesting they may feel comfortable sharing sensitive personal matters and feelings with the robot. Although previous studies have not explicitly used GenAI models to address safety issues, the evidence of children's increasing engagement with such technologies highlights significant potential for integrating them into household safety practices.

To avoid confusion, we clarify two core concepts used throughout this paper: ``GenAI'' denotes technologies like ChatGPT that generate content based on user inputs, while ``AI agent'' refers to GenAI models capable of autonomously memorizing contexts and making decisions~\cite{feuerriegel2024generative}.

We explored how families envision interacting with AI agents to support household safety. We conducted a two-phase study with 13 parent-child dyads to investigate the following questions:

\begin{itemize}
    \item[RQ1] How do parents and children conceptualize the roles of AI agents within the family context?
    \item[RQ2] What are families' preferences for AI agents' involvement in managing household safety?
    \item[RQ3] What underlying needs and concerns shape families' attitudes toward AI agent-assisted safety management?
\end{itemize}

Phase 1 consisted of individual interviews that explored families’ baseline perceptions and current experiences with GenAI. Families described these GenAI models (e.g., ChatGPT) as primarily reactive with limited autonomy, though they often anthropomorphized GenAI's interactions in their descriptions. Participants also expressed a desire for GenAI to adopt a more proactive and autonomous role in family safety protection.

Phase 2 built on these insights through co-design sessions, where parent-child dyads collaboratively envisioned how AI agents could integrate safety functionalities into everyday family routines. We found that families imagined future AI agents adopting familiar caregiving roles, such as household management, tutoring, and emotional support. Families did not seek a standalone safety AI agent, but rather anticipated that safety considerations would be naturally woven into these caregiver-like roles. The household manager's routine tasks, such as handling emails and organizing schedules, provide natural opportunities for detecting digital scams and monitoring physical safety. The private tutor's educational role easily extends to teaching driving and digital safety skills. Similarly, the family therapist's emotional support and conflict resolution responsibilities position it to address cyberbullying and sensitive issues, especially for teenagers who might want to discuss concerns with an AI agent.

In addition, our analysis uncovered important considerations regarding agent boundaries and privacy preferences. Families emphasized the need for clear boundaries between different AI agents, particularly regarding the handling of sensitive information, even though the agents are managed by the same system. We found generational differences in privacy preferences, with younger users generally showing greater openness to AI agent integration while parents expressed more privacy concerns. Both groups emphasized the importance of maintaining direct family communication alongside AI agent support, with children preferring autonomy in how safety-related information is shared with parents.

Based on these findings, we propose a multi-agent system for families in which distinct AI agents are embedded in everyday caregiving roles to support safety while maintaining agent-specific privacy boundaries. These agents operate through separate pathways to address family routines and safety needs in a way that supports direct parent-child communication. To manage information flow across these AI agents, we introduce four privacy principles: memory segregation, conversational consent, selective sharing, and progressive memory. These principles enable collaborative family data practices that protect teen autonomy and support long-term safety in everyday life.

In summary, our paper presents three key contributions:

(1) We show that families prefer multiple caregiving AI agents over a single safety-focused agent;

(2) We design a multi-agent system for families that balances parental oversight, teen autonomy, and private data sharing;

(3) We propose design guidelines for managing privacy across different AI agents.

\section{Related Work}
Our research builds upon several interconnected areas of prior work. We first examine the range of safety challenges that families face in physical and digital environments, which has motivated our investigation into AI-supported safety management. We then review existing technological solutions for family safety, highlighting their capabilities and limitations. we also explore current AI applications in family support, which provide insights into the potential and challenges of AI integration into everyday family life. Finally, we summarize how theoretical perspectives on family technology use that have guided related research and informed our study. This structure allows us to identify gaps in existing approaches and establish the foundation for our research on the future roles of AI agents in family safety management.

\subsection{Safety Issues that Teenagers Face}

Technological advances have expanded the scope of family life and brought about more safety issues, as hazards in the physical world can manifest quietly and rapidly, creating significant concerns for parents. Physical safety encompasses traffic safety~\cite{niebuhr_pedestrian_2016}, home security~\cite{r_design_2023}, and health emergencies~\cite{papachristou_is_2022}. In the physical world, families contend with various safety concerns around their homes and neighborhoods. Research~\cite{papachristou_is_2022} shows that home accidents account for 40\% of all child injuries, with health emergencies being the most common cause. For teenage pedestrians, the majority of traffic incidents occur during daily commutes to and from school or leisure activities~\cite{posner_exposure_2002}, and children are recognized as a demographic at heightened risk for traffic accidents, which frequently result in severe injuries~\cite{niebuhr_pedestrian_2016}.

Families face growing digital safety challenges as daily life becomes increasingly digitized. Children are particularly vulnerable to multiple types of digital threats, including cyberbullying, sexual exploitation, emotional manipulation, exposure to inappropriate content, privacy violations, and information security risks~\cite{freed_understanding_2023}. Safety concerns have escalated, with risks including exposure to inappropriate content, data collection without parental consent, and online predators~\cite{alshamrani_detecting_2020, yu_exploring_2024}, raising significant concerns about children's digital safety~\cite{livingstone_children_2017}. Social media platforms pose unique challenges, as Anderson et al.~\cite{gottfried_teens_2023} reported that a third of teens have been online ``almost constantly,'' increasing their exposure to harmful content and potential exploitation.

The prevalence and complexity of digital and physical safety challenges, combined with parents' limited ability to provide constant oversight, have inspired our investigation into how AI agents might support family safety management. In particular, the documented gaps in parental oversight motivated our exploration of how AI agents could serve as a supportive intermediary while respecting family dynamics.

\subsection{Technology for Solving Family's Safety-related Issues}

In recent years, there have been significant advances in technologies designed to protect family safety in multiple areas. Home monitoring systems have evolved to become increasingly sophisticated~\cite{khan_home_2024}, and modern smart security systems~\cite{r_design_2023,vasalou_doing_2025} have integrated IoT technology and AI-driven analytics to provide advanced features such as facial recognition, abnormal activity detection, and automated emergency response protocols.

The adoption of IoT devices and smart homes introduces new complexities in managing family privacy and security. As families adopt these technologies, they must balance the benefits of automation with privacy risks while managing voice assistant settings, controlling smart device data collection, and protecting personal information~\cite{yao_privacy_2019, jin_exploring_2022, sun_child_2021}. While these technologies offer safety features, their extensive data collection practices raise concerns about the security of personal information within the family~\cite{ali2020Betrayed}.

Alongside the evolution of smart homes, parental monitoring applications have also evolved to address digital safety concerns. Modern parental control apps include features such as screen time management, content filtering, and location tracking~\cite{wang_protection_2021, ghosh_matter_2018}. Research has shown that effective use of these apps can reduce children's exposure to inappropriate content~\cite{corcoran_associations_2022}. However, studies have indicated that excessive and non-adaptive monitoring can strain parent-child relationships~\cite{park_its_2024, akter_parental_2022}.

Research~\cite{livingstone_children_2017,wisniewski_parental_2017,wang_protection_2021} has revealed notable generational differences in how families understand and manage privacy and security. Parents are often focused on external threats and digital risks, while children are more concerned with protecting their privacy within the family~\cite{livingstone_children_2017}. This disconnect often leads to conflicts as children resist parental monitoring while parents struggle to balance protective supervision with children's need for independence~\cite{wisniewski_parental_2017}. Research evaluating family-oriented privacy and safety management tools has found that many existing apps rely heavily on parental controls and overlook the importance of collaborative decision-making~\cite{wang_protection_2021}. In response, recent innovations have introduced collaborative permission management systems that enable family members to make joint decisions about application access permissions~\cite{akter_co-ops_2022}.

While these technical solutions demonstrated promising capabilities, they primarily emphasized monitoring and restriction rather than support and mediation. Our work builds on these foundations by exploring how GenAI can go beyond simple monitoring to provide more relationship-focused support for family safety management.

\subsection{AI Applications in Family Support}

Prior work~\cite{sun_smart_2023, dumaru_after_2023} examined how AI has mediated family interactions beyond basic monitoring and security. Sun~\cite{sun_smart_2023} investigated how families conceptualize and negotiate AI involvement in daily activities, finding that parents and children often have different expectations and needs regarding AI assistance. Dumaru and Al-Ameen~\cite{dumaru_after_2023} also reported that AI-mediated communication can both facilitate and complicate family dynamics.

AI has shown particular promise in supporting family well-being through specialized applications~\cite{mathisen_impact_2022, shukla_smart_2023,pan_current_2020}. AI-powered nutrition and meal planning systems integrated into smart kitchen appliances help families maintain healthier eating habits and reduce food waste~\cite{mathisen_impact_2022, shukla_smart_2023}. Sleep monitoring systems using AI have demonstrated improvements in family sleep quality~\cite{pan_current_2020}.

In education, AI has revolutionized personalized learning for children learning at home. Adaptive learning platforms use AI algorithms to identify learning progress and adjust content difficulty~\cite{wang_adaptive_2019,mu_combining_2018}. Studies have shown effectiveness in using AI for personalized educational support~\cite{xu_articial_2024,zhang_mathemyths_2024}. Natural Language Processing applications enhanced reading comprehension and language learning through interactive storytelling and personalized feedback~\cite{wen_intelligent_2021, zhang_mathemyths_2024, xu_articial_2024}. Prior work has also suggested that AI can collect and analyze multi-dimensional data from family members' daily activities, potentially serving as an early warning system for various issues. Nonetheless, AI's integration into family life requires careful consideration of both technical capabilities and social implications~\cite{ackerman_intellectual_2000}.

The diverse applications of AI in family support, along with their documented benefits and challenges, have provided the foundation for our research into AI-assisted family safety management. Our work extends these findings by specifically examining how AI can integrate safety considerations into everyday family interactions while maintaining appropriate privacy boundaries and supporting positive family dynamics.

\subsection{Theoretical Perspectives on Technology Use in Families}

To understand how families adopt and manage safety-related technologies, we utilized three theoretical perspectives: 
Family Systems Theory (FST)~\cite{broderick1993understanding}, Communication Privacy Management (CPM) theory~\cite{petronio_communication_2010}, and existing research on technology adoption in family contexts~\cite{davis1989perceived,brown2005model,de2021exploring,de2019would}. These frameworks illuminated how technology reshaped family roles, guided privacy boundary management within families, and affected family decisions around adoption. They provided us with key perspectives to study the future integration of AI agents into household safety practices.

FST~\cite{broderick1993understanding,cox1997families} views the family as an interconnected system where new technology can transform established routines and relationships. Studies~\cite{cagiltay2023family,beneteau2020parenting} applying FST to smart homes and social robots demonstrated how technologies became integrated into family dynamics as more than tools, influencing caregiver roles and emotional connections. This theory has suggested that safety technologies should be designed sensitively to complement existing family dynamics.

In addition, theoretical research~\cite{davis1989perceived,brown2005model} on technology use in families emphasized that decisions to adopt new tools were influenced not only by perceived usefulness or ease of use but also by emotional, relational, and hedonic motivations. For instance, children and parents often hold different preferences and expectations for digital tools, exposing gaps in the design of safety technologies~\cite{de2021exploring,de2019would}. These findings underscored the importance that technology for family use should account for diverse values and user experiences across generations.

Finally, CPM theory~\cite{petronio_communication_2010} addresses how families manage private information. Research~\cite{vitak2023boundary,moser2017parents} using CPM theory in smart home contexts and parent-child digital monitoring highlights tensions over personal data control, especially when safety intersects with autonomy. As we study how AI agents may integrate into daily family interactions, CPM provides us insights into how to design AI agents to better manage trust and privacy effectively.

These theoretical frameworks help explain the challenges families encounter when adopting and using home security technologies, guiding our research on how AI agents can better address gaps in family safety needs and expectations.

\section{Methodology Overview}
\begin{figure}[htbp]
  \centering
  \includegraphics[width=\linewidth]{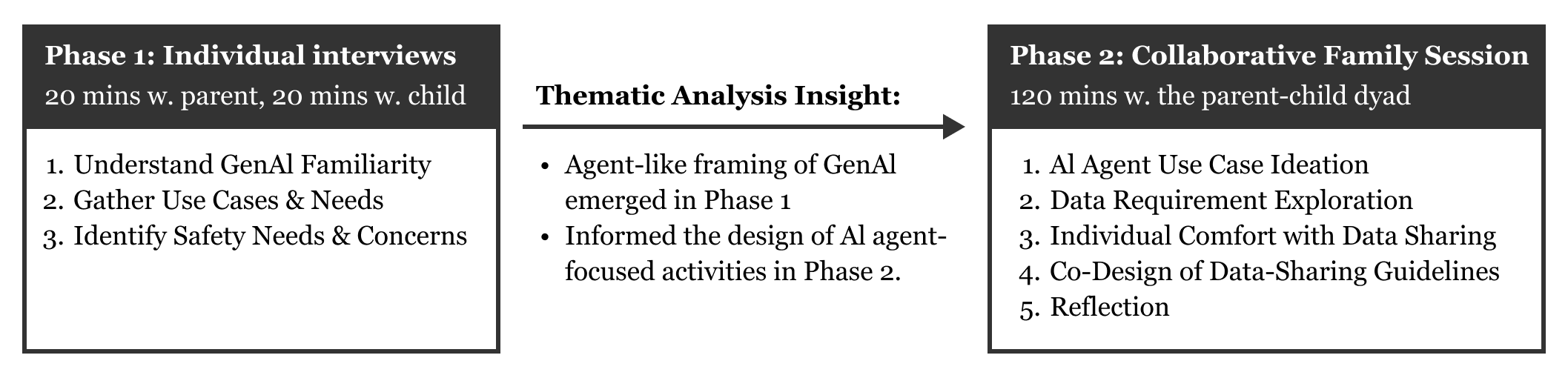}
  \caption{Two-Phase Study Method Overview. Phase 1: separate 20-minute individual interviews with the parent and then with the child on GenAI familiarity, use cases, and safety expectations. Phase 2: 120-minute family sessions on AI agent ideation, safety-related data-sharing, privacy boundary, and reflection.}
  \label{fig:method-flowchart}
  \Description{A two-phase family study design: Phase 1 uses short parent and child interviews to understand GenAI familiarity, needs, and safety concerns. Insights about agent-like views of GenAI inform Phase 2, a longer joint session where families ideate AI use cases for household safety.}
\end{figure}

We conducted a two-phase qualitative study to explore how families envision GenAI's roles in managing household safety, focusing on families with teenagers (ages 13 to 15). The university IRB reviewed and approved this study. The study process is illustrated in Figure~\ref{fig:method-flowchart}.

In Phase 1 (details in Section~\ref{sec:method-p1}), we conducted individual interviews with family members to understand their baseline perceptions, uses, and concerns around GenAI. 

In Phase 2 (details in Section~\ref{sec:method-p2}), each family returned one week later for a two-hour collaborative session. The session design drew on a perspective-taking protocol for structuring intergenerational discussions about digital privacy~\cite{wen2025supporting}, which we adapted to the context of AI-enabled household safety to elicit diverse viewpoints and support boundary negotiation. At the start, we introduced participants to future-oriented GenAI with agentic capabilities, such as remembering context, learning from interactions, and autonomously coordinating tasks. Participants then discussed potential AI agents for household safety, as well as family data-sharing preferences and privacy boundaries when AI agents come into play. 

\subsection{Participants}

We recruited 13 parent-child dyads from local communities in the United States through flyers and social media advertisements. Children were between 13 and 15 years old (M = 14.38 years, SD = 0.87). This age range was chosen because we would like the families to discuss safety issues related to social media platforms that required users to be at least 13 years old to register. Parents were between 35 and 64 years old.

Table~\ref{tab:userInfo} provides detailed information on the parents and children in our study. Our sample included a variety of work status backgrounds, with most parents working full-time (9 of 13) or part-time (3 of 13), and one self-employed parent. Guardians had a variety of educational levels, with most having completed at least a bachelor's degree. The racial composition of the sample was primarily Caucasian (10 of 13) and Asian (3 of 13). Family size ranged from 2 to 11 members (M = 4.38, SD = 2.22). Of the child participants, seven were male, five were female, and one was non-binary. The children were in grades 7 to 11, with the majority in grades 9 and 10.

\begin{table}[htbp]
\centering
\footnotesize
\caption{Study Participant Demographic Information: The ID starting with an `F' refers to the dyad of parent and child. Letter `P' refers to the parent of the family, and letter `C' refers to the child.}
\Description{This table summarizes the demographics of the 13 families in the user study. It includes guardian information (age, role, race, job, education), family size, and child details (age, gender, grade). Guardians are mostly mothers aged 35-54, predominantly white, with full-time employment and higher education. Children are aged 13-15, in grades 7-11, and are a mix of both sexes. Family size ranges from 2-11 people.}
\label{tab:userInfo}
\begin{tabular}{rrccccrrcr}
\toprule
\multirow{2}{*}{ID} & \multirow{2}{*}{Age} & Guardian & \multirow{2}{*}{Ethnicity} & Work & Guardian & Family & Child & Child & \multirow{2}{*}{Grade} \\
 & & Role & & Status & Education & Size & Age & Gender & \\
\midrule
F01  & 35-44 & Mother   & Asian         & Full-time      & Graduate      & 4           & 13        & Male   & 7     \\
F02  & 35-44 & Mother   & Asian         & Full-time      & Graduate      & 4           & 15        & Male   & 9     \\
F03  & 35-44 & Mother   & Caucasian     & Part-time      & Bachelor      & 5           & 15        & Male   & 9     \\
F04  & 45-54 & Mother   & Caucasian     & Part-time      & Graduate      & 4           & 15        & Female & 9     \\
F05  & 45-54 & Father   & Caucasian     & Full-time      & Graduate      & 4           & 15        & Female & 10    \\
F06  & 35-44 & Father   & Caucasian     & Full-time      & Bachelor      & 6           & 13        & Female & 9     \\
F07  & 45-54 & Mother   & Caucasian     & Full-time      & Graduate      & 2           & 13        & Female & 9     \\
F08  & 45-54 & Father   & Caucasian     & Self-employed  & Bachelor      & 11          & 15        & Female & 10    \\
F09  & 45-54 & Mother   & Caucasian     & Part-time      & Graduate      & 4           & 15        & Male   & 10    \\
F10  & 35-44 & Mother   & Caucasian     & Full-time      & College       & 4           & 15        & Male   & 9     \\
F11  & 35-44 & Mother   & Asian         & Full-time      & Graduate      & 3           & 14        & Male   & 9     \\
F12  & 45-54 & Mother   & Caucasian     & Full-time      & Bachelor      & 3           & 15        & Female & 11    \\
F13  & 55-64 & Mother   & Caucasian     & Full-time      & Bachelor      & 3           & 15        & Non-binary & 10    \\
\bottomrule
\end{tabular}
\end{table}

Each parent and each child received \$20 per hour, paid via Amazon gift cards. If families completed the study within two hours, they were still paid for two hours.

\subsection{Consent Process}
We designed our informed consent process so that it aligned with best practices for family-based research involving minors~\cite{voltelen_ethical_2018}. Before participating, parents and children received separate explanations about the study, tailored appropriately for each age group. For children, we explicitly stated their right to decline participation independently of their parents' consent. We informed all families that compensation would be provided regardless of whether they completed the entire study, ensuring participation remained voluntary. We also explained that although complete anonymity among family members could not be guaranteed in our study, we implemented password-protected measures to protect the confidentiality of all collected personal data.

\subsection{Data Collection and Analysis}

All sessions were audio-recorded with participant consent. Recordings were transcribed. Additional data sources included researcher field notes, participant-generated artifacts from workshop activities, and post-session reflection documents. The researchers then followed Saldana’s approach~\cite{saldana2021coding} to qualitatively code the transcripts using a constant comparative method to identify common themes across families.

The coding process was iterative. Initially, two researchers independently coded three sample transcripts and then discussed and agreed on codes and themes. One researcher coded the remaining transcripts according to these agreed-upon categories. After drafting an initial report and reflecting on the findings, we repeated the coding process. In the rare cases of disagreement between coders, discussions were held until consensus was reached. One researcher then led the recoding of the remaining transcripts, and the other researcher reviewed and discussed any consistent or inconsistent points with the lead coder.

Our approach follows the coding method used in prior research, such as identification of ``folk models'' to understand home computer users' perceptions of security threats by Wash~\cite{wash2010folk}, and structuring of social cybersecurity literature into key behavioral domains by Wu et al.~\cite{wu2022sok}. By employing the iterative and collaborative process, we aimed to ensure that our code organization and domain generation were both systematic and reflective of the data, thereby enhancing the reliability of our findings.

\section{Phase 1: Individual Interviews on GenAI Experiences and Safety Perceptions}
\subsection{Study Procedure}
\label{sec:method-p1}

We conducted separate 20-minute semi-structured interviews with each parent and each child to establish a baseline understanding of GenAI familiarity and usage patterns. 

We began by explaining to participants that GenAI refers to AI technologies capable of generating new content, suggestions, or interactions based on user prompts (e.g., ChatGPT). We did not limit participants to an agent-based notion. We clarified that such technologies generate content probabilistically using large datasets collected from humans, which may produce incorrect or inaccurate information, as well as potentially biased or inappropriate content.

Next, we introduced a popular GenAI model, ChatGPT, and demonstrated how participants could use it in three interaction modes (i.e., text, voice, image). After a hands-on interaction, participants evaluated their comprehension across different interaction modes, assessed the perceived effectiveness of ChatGPT, and provided rationales for their evaluation. We then asked participants if they had used other partial or complete GenAI models at home (e.g., additional chatbots, voice assistants, or email editors) and if they held any safety concerns regarding GenAI.

The detailed scripts of the study procedure, along with explanations of generative AI terms and the interview questions, are in Appendix~\ref{app:interview-script}.
\subsection{Findings}

The Phase 1 interview analysis revealed varied engagements with GenAI models among parents and children, which we categorized into four themes: educational support, everyday guidance, entertainment interactions, and emerging autonomous safety management. Although most current uses of GenAI models involved reactive responses to explicit user commands, participants' discussions about household safety and digital monitoring hinted at a shift toward imagining more proactive and context-aware AI agents.

Participants used GenAI models like ChatGPT for educational tasks, emphasizing its practical value in information retrieval and assistance. For instance, C03 (age 15, male) described using GenAI to \textit{``rewrite my stories or homework, so it becomes easier to understand,''} underscoring the tangible support provided in academic contexts. Similarly, P01 (age 35–44, mother) highlighted GenAI's real-time responses as advantageous for educational purposes.

Parents also valued GenAI models for their everyday guidance and practical support. They frequently relied on these tools for navigation assistance, quick information retrieval, and general household management tasks. For example, P01 (age 35–44, mother) specifically mentioned talking to Google Maps to find and navigate to facilities such as family restrooms during outings. Additionally, P09 (age 45–54, mother) appreciated ChatGPT's suggestions for effective communication with her child, helping bridge conversational gaps such as interpreting teen slang.

Entertainment emerged as another central domain, especially among children, who often engaged with GenAI casually for amusement and conversational interactions. For example, C03 (age 15, male) praised ChatGPT's conversational quality, stating it \textit{``feels like I'm really talking to a friend.''} Other children shared similar sentiments. C04 (age 15, female) noted that her chosen voice assistant sounded genuinely empathetic, though somewhat robotic, expressing a desire for GenAI to behave less like a comforting human presence and instead have its own thoughts. Meanwhile, C06 (age 13, female) described using ChatGPT to generate whimsical and surreal images, such as a cat riding a horse through space, delighting in ChatGPT's playful responsiveness. These interactions highlight children's playful and emotionally resonant engagement with GenAI.

Notably, participants in the interviews showed minimal concern about GenAI itself posing safety risks; instead, parents expressed substantial interest in employing GenAI for managing household and digital safety. They often described GenAI's potential roles in activity monitoring, mediating sensitive conversations, and safely guiding children’s online interactions. P09 (age 45–54, mother) particularly emphasized GenAI's potential in managing online harassment, describing how GenAI could support her child during challenging conversations in online gaming, reminding her child that \textit{``you don't have to stick around if you are harassed.''}

Overall, participants primarily used GenAI as responsive tools that follow explicit commands. Nevertheless, their perspectives on household safety pointed toward a future vision of AI agents that can act proactively. Parents imagined GenAI independently facilitating parent-child dialogues and overseeing children's online interactions, while children showed openness toward GenAI working with them like a friend. Their insights helped us revise the design of Phase 2 activities, facilitating family exploration into how future AI agents might address household digital and physical safety threats.

\section{Phase 2: Family Sessions on AI Agent Safety Roles and Privacy Discussion}
\subsection{Study Procedure}
\label{sec:method-p2}

One week after the interviews, each family returned for a two-hour collaborative session. During the session, we facilitated a series of activities (A2.1--A2.5). Participants first brainstormed potential future uses of AI agents (A2.1) and then considered the data requirements for their ideas (A2.2). Next, they rated their individual comfort levels with sharing such data with AI agents and with other family members (A2.3), followed by a group discussion to co-design family-level data-sharing guidelines (A2.4). Finally, they reflected on any evolving perspectives around AI agent adoption and safety (A2.5). This collaborative process enabled both parents and children to consider how AI agents might be integrated into their routines while balancing convenience, privacy, and safety. The details of each activity are as follows:

\begin{enumerate}
    \item[\textit{A2.1}] \textit{AI Agent Use Case Ideation (45 minutes):} We began by presenting the four categories identified from participants' responses in Phase 1: education, guidance and support, entertainment, and future safety integration. Next, we summarized participants' ideas about GenAI's future agentic capabilities, including remembering context, learning from interactions, and coordinating tasks autonomously. Following this summary, each family member individually wrote down on whiteboards how they imagined future AI agents could support their households. We required them to adopt the other person's perspective, with parents imagining from their child's viewpoint and children from their parent's viewpoint. We also encouraged them to think beyond their current experiences with GenAI and consider how GenAI might develop in the next few years. Afterwards, participants came together to share their ideas and discuss how AI agents might fit into different household routines.
    \item[\textit{A2.2}] \textit{Data Requirement Exploration (20 minutes):} Participants continued to work as a group to map out what types of personal and family data would be needed to enable each proposed AI agent idea. This activity helped them weigh the convenience and safety benefits against potential privacy trade-offs.
    \item[\textit{A2.3}] \textit{Individual Comfort with Data Sharing (15 minutes):} Each participant rated their own comfort level with sharing the data types identified above, focusing on potential impacts on family safety. They completed the ratings individually to minimize immediate peer influence.
    \item[\textit{A2.4}] \textit{Family Co-Design of Data-Sharing Guidelines (25 minutes):} Participants then came together again to work as a group in a facilitated session to co-design how AI agents could access and handle family data. They discussed potential privacy-safety trade-offs, how AI agents might mediate or filter family information flow, and identified conditions under which sharing specific data would feel acceptable. We encouraged both parent and child to propose design solutions that balanced each person's comfort with family-level needs.
    \item[\textit{A2.5}] \textit{Reflection (5 minutes):} Participants reflected on insights gained in the study. They discussed how their views on GenAI evolved through the activities and formulated guidelines for future AI agent adoption that would preserve safety and privacy within the household. we provided clarifications if participants expressed misconceptions or confusion about AI agents. We also asked if they had concerns or discomfort at any point, assuring them we would work with them now or after the study until we address all the issues (if any).
\end{enumerate}

\subsection{Findings Overview}

We present our findings in three subsections. Firstly, we describe how families envision AI agents in everyday roles (RQ1, Section~\ref{sec:roles}). Next, we connect these roles to the specific safety threats families anticipate (RQ2, Section~\ref{sec:safety-findings}). Finally, we highlight underlying design needs and broader concerns influencing families' willingness to adopt AI agents for household safety (RQ3, Section~\ref{sec:concerns}).

\subsection{Roles of AI Agents within the Family Context (RQ1)}
\label{sec:roles}
Parents and children made sense of AI agents by personifying them as caregivers in the family, which helped them articulate desired functionalities and interaction patterns. While participants were aware that GenAI might provide inaccurate or incorrect information, they nonetheless envisioned it taking on helpful and autonomous caregiving agents. We categorized their expectations into three agents: (1) a household manager for day-to-day tasks, (2) a private tutor for educational support, and (3) a family therapist for mediating relationships and providing emotional support.

\subsubsection{The Household Manager Agent}

Participants envisioned the AI agent as a sophisticated household manager that filters digital communications, manages household schedules, organizes chores, and recommends entertainment.

When discussing potential AI agent assistance with daily tasks, C05 (age 15, female) expressed frustration that her father spent too much time on emails instead of family time, saying:
\begin{quote}
    ``AI should help with processing emails because you spent so much time on emails, and I think that if you had something to help you with your emails, you might be able to do some things, such as like hanging out with us more.''
\end{quote}
P05 (age 45-54, father) laughed upon hearing his daughter's suggestion, agreeing that ChatGPT seemed more capable of replying on his behalf compared to auto-complete, but added that it was not quite there yet. He then extended their conversation to his difficulty identifying phishing emails, suggesting that a future AI agent alert him which emails are phishing emails and explain why.

Similarly, P09 (age 45-54, mother), whose work involves responding to client emails, felt an AI agent could screen suspicious emails so she would not \textit{``waste time weeding out potential scams.''} C09 (age 15, male) nodded in agreement.

Families also mentioned daily routines that current methods handled poorly, wondering if an AI agent could do better. P06 (age 35-44, father) and C06 (age 13, female) discussed:
\begin{quote}
    \enquote{Make chore lists... keep track of who's done what, so we don't argue about like \enquote{last time it's your turn!}}
\end{quote}

C10 (age 15, male) provided another example, expressing his hope: \textit{``I cannot wake up for the life of me... AI can push me like `[C10's name], it's time to get up in the morning'.''} Since both C10 and P10 (age 35-44, mother) felt P10's wake-up attempts were not that effective and imagined an AI agent might have greater patience nudging C10 awake, reducing their morning frustrations.

\subsubsection{The Private Tutor Agent}

Children particularly valued the AI agent's potential to provide on-demand tutoring support when traditional teaching methods fall short. C08 (age 15, female) said, \textit{``School's not really working out for me as well, and I've been kind of getting a lot of homework, and it's like kind of hard. Like with geometry... my teacher doesn't really explain it that well.''} Others, like C07 (age 13, female), also discussed with their parents that ChatGPT could evolve into a tutor \textit{``who can explain things in more than one way''} and adapt to their own learning pace.

Parents saw AI agent's academic support expanding into digital safety literacy and broader safety skills training. For instance, P06 (age 35–44, father) envisioned the tutor teaching critical thinking to detect fake news when his child scrolling social media, adding:
\begin{quote}
    \enquote{It's getting harder and harder to tell what's real and what's not out in the news world. So that [AI agent] can be a useful teacher on, \enquote{Hey, I heard this story...}}
\end{quote}

Although we did not observe children proactively requesting tutoring in digital safety literacy or broader safety skills before their parents mentioned it, they agreed with their parents' ideas, as it provided \textit{``another person''} to learn from besides their parents.

\subsubsection{The Family Therapist Agent}

Families also envisioned the AI agent acting as a family therapist who could mediate relationships, support emotional well-being, and address sensitive topics that might otherwise be difficult for children to tell their parents. Participants appreciated the AI agent's potential for improving family communication. 

For example, P13 (age 55-64, mother) described it as \textit{``a generational translator''} to help bridge misunderstandings with her child:
\begin{quote}
    ``Well, kind of like Gen. Z and Gen. X translator. A translator, maybe like a... slang translator. I'm thinking of the AI like someone pulling out old letters from the attic from your great, great granddad and trying to understand what those words mean.''
\end{quote}

Children saw the AI agent as helpful for resolving conflicts with siblings. For example, C05 (age 15, female) imagined the AI agent analyzing past sibling disputes, explained, \textit{``If I previously had an argument with my brother, how would I deal with it? Like how did I deal with it before?''} She felt the AI agent could offer objective suggestions or helpful reminders to reduce tension at home so her father would not have to take sides in their conflicts.

Another aspect of the therapist's role involved supporting children's emotional well-being by providing a safe outlet for sensitive conversations that children might feel hesitant to discuss with parents. Parents viewed the AI agent as an additional supporter with whom their children could share personal struggles. As P03 (age 35-44, mother) explained:
\begin{quote}
    ``When it has a screening for mental health, I think that would be helpful because there's just a lot about my boy's inner world that I won't know.''
\end{quote}

Children expressed varied reactions to this kind of idea suggested by their parents. Some teenagers valued an AI listener. C04 (age 15, female) highlighted this need:
\begin{quote}
    ``AI would be great because you can talk to the AI every day... I don't specifically talk to my parents about messages [with friends] sometimes because it's sometimes difficult to talk to my parents about it.''
\end{quote}

A few children strongly rejected this suggestion by their parents, such as C01 (age 13, male), who dismissed this idea outright, responded to his mother, \textit{``A talking buddy like [ChatGPT] is stupid. I don't need a talking buddy.''} This variation suggests that teenagers differed in their comfort with forming close relationships involving an AI therapist.

\subsection{Integration of Safety Features Across Family Roles (RQ2)}
\label{sec:safety-findings}

After identifying three distinct caregiving agents envisioned by families, we report how each agent integrates safety features into its specialized domain. We first describe how families naturally discussed nine digital and physical safety threats when imagining AI agents performing everyday tasks. Then, we map each threat to its corresponding AI agent.

\subsubsection{Identification and Contextual Emergence of Safety Threats}

We identified 26 discussion excerpts about safety threats in everyday family life and then clustered them into six digital threats (sexual violence, harassment/cyberbullying, financial fraud, digital addiction, unsafe or illegal behaviors, and misinformation/deepfakes) and three physical threats (unsafe driving, home safety incidents, and health emergencies).

Notably, these safety concerns arose while families discussed how AI agents could assist with daily tasks, such as email management, academic learning, and social media, rather than as standalone safety issues. For example, in a conversation about how an AI agent might help reply to mobile messages when they were busy, C04 (age 15, female) commented, \textit{``I don't really feel unsafe on the Internet, but I do feel unsafe sometimes if like a random account follows me or DMs me.''} P04 (age 45–54, mother) recommended blocking or reporting those accounts, then asked, \textit{``What information could we give AI to help protect you online?''} C04 then reflected on what data she might provide to help the AI agent recognize suspicious interactions when assisting with texting, such as someone from a former school contacting her.

Therefore, our analysis contextually maps safety issues to corresponding AI agents. For instance, when families discussed daily tasks typically performed by a household manager and then mentioned relevant safety concerns and how to address them, we mapped these functions to that agent, even if participants did not explicitly name it.

\subsubsection{Embedding Safety Features within Daily Family Roles}
\label{sec:safety-features}

In this subsection, we summarize how the household manager, private tutor, and family therapist agents integrate safety features into everyday family routines. Table~\ref{tab:safety-features} illustrates how the routine tasks performed by each agent address the nine identified safety concerns.

\begin{table}[htbp]
    \centering
    \caption{Multiple AI Agents and Their Integration of Family Safety Needs into Everyday Tasks}
    \renewcommand{\arraystretch}{1.1} 
    \setlength{\tabcolsep}{10pt} 
    \footnotesize
\begin{tabular}{>{\raggedright\arraybackslash}p{0.1\textwidth}>{\raggedright\arraybackslash}p{0.08\textwidth}>{\raggedright\arraybackslash}p{0.09\textwidth}>{\raggedright\arraybackslash}p{0.12\textwidth}>{\raggedright\arraybackslash}p{0.23\textwidth}>{\raggedright\arraybackslash}p{0.08\textwidth}}
        \rowcolor{headercolor}
        \textcolor{white}{\textbf{AI Agent}} & \textcolor{white}{\textbf{Threat Type}} & \textcolor{white}{\textbf{Threat Subtype}} & \textcolor{white}{\textbf{Risk Indicator}} & \textcolor{white}{\textbf{Safety Management Approach}} & \textcolor{white}{\textbf{ID}}  \\
        \toprule
        
        \textbf{Household Manager} & Digital & Financial Fraud & Scams & Processes incoming communications for scams; Verifies transactions and sellers information & F06, F09  \\
        \cmidrule(lr){3-6}
        & & Unsafe, Illegal Behaviors & Underage Alcohol Purchase & Monitors and alerts parents about concerning purchases & F11 \\
        \cmidrule(lr){3-6}
        & & Digital Addiction & Excessive Screen Time & Manages screen time and encourages healthy habits & F02 \\
        \cmidrule(lr){2-6}
        & Physical & Home Safety Incident & Appliance Safety & Monitors home environment and eliminates safety risks &F02, F04, F10 \\
        \cmidrule(lr){3-6}
        & & Health Emergency & Medical Conditions, Sudden Illness or Injury & Monitors health indicators and manages medications & F04, F10, F11, F12, F13 \\
        \midrule
    
        \textbf{Private Tutor} & Digital & Mis-information, Deepfakes & Fake News Exposure & Teaches critical thinking skills to protect from misinformation & F06 \\
        \cmidrule(lr){2-6}
        & Physical & Unsafe Driving & Unsafe Driving & Provides driving safety education & F01, F06, F10, F12 \\
        \midrule
        
        \textbf{Family Therapist} & Digital & Sexual Violence & Non-consensual Imagery, Explicit Content & Provides private space for discussing sensitive content concerns & F03, F05, F08 \\
        \cmidrule(lr){3-6}
        & & Harassment and Cyberbullying & Toxic, Reputation Damage, or Abusive Content  & Monitors online communication patterns for signs of bullying and provides emotional support & F01, F02, F05, F06, F07, F08 \\
        \bottomrule
    \end{tabular}
    \label{tab:safety-features}
\end{table}

Participants envisioned the \textbf{AI household manager} embedding safety naturally into everyday tasks like screening emails for scams, managing screen time, and quietly monitoring online behaviors or household hazards.

Families F06 and F09 envisioned the AI agent seamlessly integrating scam detection into daily communication and financial tasks. Similarly, family F11 proposed that the AI agent could alert P11 (age 35-44, mother) to unsafe online activities, such as her child buying beers using her online shopping account. Additionally, family F02 connected the AI agent's daily scheduling functions with preventing digital addiction by regulating screen time. P02 (age 35–44, mother) imagined a fun intervention from the AI agent, suggesting it say, \textit{``Buddy, you've been on your device for far too long. Get up and get out!''} C02 (age 15, male) echoed this sentiment, acknowledging, \textit{``Sometimes I just don't have anything to do,''} and suggested that the AI agent could, in addition to reminders, propose alternative activities such as recommending interesting outdoor events to enrich his free time.

In the physical world, participants viewed the AI agent's ongoing monitoring as naturally suited to managing everyday household hazards. C02 (age 15, male) emphasized practical scenarios, noting the AI agent could notice if users \textit{``forget to lock your doors...or forget to turn off something like the stove.''} Additionally, participants highlighted the agent's potential in responding to health emergencies. C04 (age 15, female) wondered if the AI agent could assist in urgent health situations, such as, \textit{``Like my friend coughs up blood...we can ask AI if it's OK that I'm coughing up blood?''} Complementing her concern, P04 (age 45-54, mother) envisioned the AI agent rapidly assessing the emergency context, such as checking for \textit{``a bottle of pills''} near an unconscious teenager.

Participants envisioned the \textbf{AI private tutor} naturally extending everyday academic support to include safety education, making the learning of safety skills feel organic rather than rule-driven. For instance, P06 (age 35–44, father) described how the AI agent could incorporate digital literacy by teaching critical thinking skills to help teenagers discern credible information online. It could also guide teenagers in preparing for safe driving by suggesting relevant topics and practical tips as they approach driving age.

Participants envisioned the \textbf{AI family therapist} extending its everyday emotional support role to naturally include handling sensitive, safety-related conversations. When addressing sexual violence, P03 (age 35–44, mother) described how teenagers might feel more comfortable approaching an AI agent by explaining: 
\begin{quote}
    ``What to do if someone sends you pictures you don't want to see... instead of not being sure how to bring it up [to me], he could just like go to AI.''
\end{quote}
Similarly, C05 (age 15, female) imagined proactive interventions from the AI agent, suggesting it could prompt teenagers by asking, \textit{``If you send a sexually explicit photo on like your iPhone... It'll tell you like you sure you wanna send this?''}

Participants also described how the therapeutic relationship provided a natural context for addressing cyberbullying and online harassment. P02 (age 35–44, mother) envisioned the AI agent as an unobtrusive mediator, saying, \textit{\enquote{As if AI is that silent person in that conversation... and basically says, \enquote{This is not OK,} or \enquote{This seems like you have a mental issue you would like to talk about?}}}

\subsection{Underlying Needs and Concerns Influencing Families' Attitudes (RQ3)}
\label{sec:concerns}
Building on the ways families integrate safety features into daily routines, we now report the broader issues that shape their readiness to adopt AI agents. Specifically, we highlight three key themes: concerns about privacy management, generational differences in trust, and the desire to preserve in-person family communication.

\subsubsection{AI Agent Privacy Boundaries and Private Information Management}

Participants emphasized the need for clear boundaries between different agents. For instance, the household manager agent within the system should not automatically share data with other agents, such as the private tutor or family therapist, even though all are managed by the same underlying system. As P05 (age 45–54, father) explained:
\begin{quote}
    ``Maybe is the information that another one has the right to keep private, and so just like because therapist knows it doesn't mean necessarily that others have the same privilege of knowing it.''
\end{quote}

Additionally, participants expressed worries about potential external misuse or leaks of personal information. For instance, P11 (age 35–44, mother) expressed concern about the risk if AI agents were to inadvertently expose their family's financial or travel details, emphasizing the serious consequences if such sensitive data fell into malicious hands. C11 (age 14, male) agreed and described this as \textit{``a cybersecurity hazard.''}

Overall, families agreed that clearly defined privacy boundaries among AI agents were essential for maintaining trust, avoiding confusion, and preventing potential misuse or external leaks of sensitive information.

\subsubsection{Generational Differences in Trust and Privacy Preferences toward AI Agents}

Building upon these role boundaries, our analysis revealed distinct generational perspectives on trust and privacy toward AI agents. Teen users generally showed greater openness, viewing it as a trusted friend. This trust stemmed partly from the perceived non-judgmental nature of AI agents, as C01 (age 13, male) stated, \textit{``AI won't judge me. Whatever happened.''}

In contrast, parents expressed more concerns about privacy and information misuse. P13 (age 55-64, mother) noted:
\begin{quote}
    ``I think it's creepy how advanced AI is... it's invasive. I can see this usefulness, but it also seems kind of invasive in a lot of ways.''
\end{quote}

Parents stressed that AI agents should not collect all possible data and automatically process the data just to complete tasks. Under this premise, they further pointed out that these AI agents should work together with family members in a way that enables users to understand and clearly express how customizable data-sharing protocols are established to accommodate different family members' varying needs for privacy, use cases, and functionality.

\subsubsection{Balancing AI Agent Integration with Direct Family Communication}

As families warmed to the idea of integrating AI agents into household routines, both parents and children emphasized the need to maintain direct communication about safety issues. Children expressed a desire for autonomy to first discuss safety issues with AI agents and then discuss the issues with their parents using their own terms. Parents also prioritize face-to-face dialogue with their children to foster family trust. As P05 (age 45-54, father) put it,\textit{ ``In the end it should be that [C05's name] and I are communicating [about the safety issues] not through the AI.''}

Participants, therefore, framed AI agents' safety support as strengthening core family conversations, rather than replacing. While acknowledging agents might offer convenient or less judgmental channels, they wanted to communicate and make the final decisions about safety protection themselves.

\section{Discussion}
In this section, we reflect on how families' visions of AI agents, expressed through everyday caregiving roles and integrated safety expectations, inform the design of a multi-agent system for families. Families envisioned safety as part of different caregiving agents rather than a standalone feature. Drawing on these expectations and theoretical insights, we propose a multi-agent system design that supports family safety and privacy by distributing safety functionalities across specialized agents, managing privacy through informed consent and selective data sharing, and enhancing direct family interactions around sensitive issues.

\subsection{User Expectations for AI Agent Safety Solutions}

\subsubsection{Envisioning Integrated Safety Approaches}

Families increasingly felt overwhelmed by managing separate safety systems for digital and physical threats. Parents expressed frustration with juggling various monitoring tools~\cite{ghosh_matter_2018, modecki_what_2022}, while children felt burdened by intrusive restrictions~\cite{wisniewski_parental_2017, park_its_2024}. This fragmentation not only increases stress but also disrupts natural family dynamics.

In our study, families did not see safety support as provided by a designated safety AI agent. Instead, they preferred embedding safety features into different caregiving agents discussed earlier, allowing AI agents to support safety seamlessly through existing family routines. 

For example, when an AI household manager assists with daily tasks like email management, it can seamlessly extend to screening for potential threats. Embedding safety functions within familiar roles makes the AI agent's actions feel natural rather than intrusive. This approach reduces the cognitive burden on family members and enhances their acceptance of safety measures. Families' preference echoes the concept of ``calm technology''~\cite{case_calm_2015}, where technology becomes a seamless part of the user's environment without demanding excessive attention. 

These role-based conceptualizations not only align with Family Systems Theory~\cite{broderick1993understanding,cox1997families}, which frames technology as influencing and reshaping familial roles and routines but also extend it by showing how AI agents can be perceived as taking on relational caregiver roles rather than remaining a functional tool. By embedding safety within these roles, families preserved the integrity of existing family dynamics while welcoming technological augmentation.

\subsubsection{Privacy and Trust in Multiple AI Agents}

Our study reveals complex trust dynamics between family members and AI agents. Parents like P13 described the system as \textit{``creepy and invasive,''} citing incomplete transparency in data handling, while teenagers like C01 trusted GenAI for its non-judgmental responses, stating, \textit{``AI won't judge me. Whatever happened.''} Although teenagers were aware that GenAI's responses were probabilistically generated rather than genuinely empathetic, they still perceived less judgment than from peers or parents. This generational differences in prioritizing non-judgmental interactions versus protecting personal secrets highlights the challenge of designing AI agents that accommodate diverse trust needs within the family.

Agent-specific trust relationships further complicate privacy management. Children might share sensitive information with the AI therapist but not with the AI tutor or their parents. Parents desire oversight but also want to respect children's autonomy, making the design of a scheme for private information-sharing necessary.

CPM theory~\cite{petronio_communication_2010} explains family privacy boundaries, and our findings extend this theory by showing how these boundaries must also be negotiated \textit{within} a multi-agent AI system serving multiple relational agents. Families emphasized that information shared with one AI agent should not be transferred to another, emphasizing the need for agent-specific boundary coordination not previously addressed in CPM.

A lack of guidelines for natural consent mechanisms and flexible privacy controls for cross-agent information sharing can undermine trust and discourage full engagement. To address this gap, we propose four design principles (i.e., memory segregation, conversational consent, selective sharing, and progressive memory) that build on CPM's emphasis on negotiated disclosure, adapted to a multi-agent family context (details in Section~\ref{sec:cpm-extend}).

\subsubsection{Enhancing Instead of Replacing Family Safety Communication}

Children expressed a desire for autonomy in handling sensitive issues, preferring to discuss concerns with AI agents before involving their parents. Children's proposal offers them time to prepare for challenging conversations and maintain greater control over their information. Parents emphasized the importance of direct communication but also welcomed this idea.

This idea reflects prior theoretical research on family-centered technology adoption~\cite{de2021exploring,de2019would}, which links adoption to perceived usefulness and social influence. However, our findings suggest that trust in AI agents is shaped not only by utility but also by the relational tone and perceived emotional alignment of the agent. Children's willingness to disclose their sensitive thoughts with an AI therapist stemmed from a sense of non-judgmental interaction rather than just functional benefit, indicating that emotional safety is a critical but underexplored factor in such AI technology adoption within families.

\subsection{Multi-Agent System Design for Family-Centered Safety and Privacy}

\subsubsection{Family-Centered Multi-Agent System Design}

\begin{figure}[htbp]
  \centering
  \includegraphics[width=\linewidth]{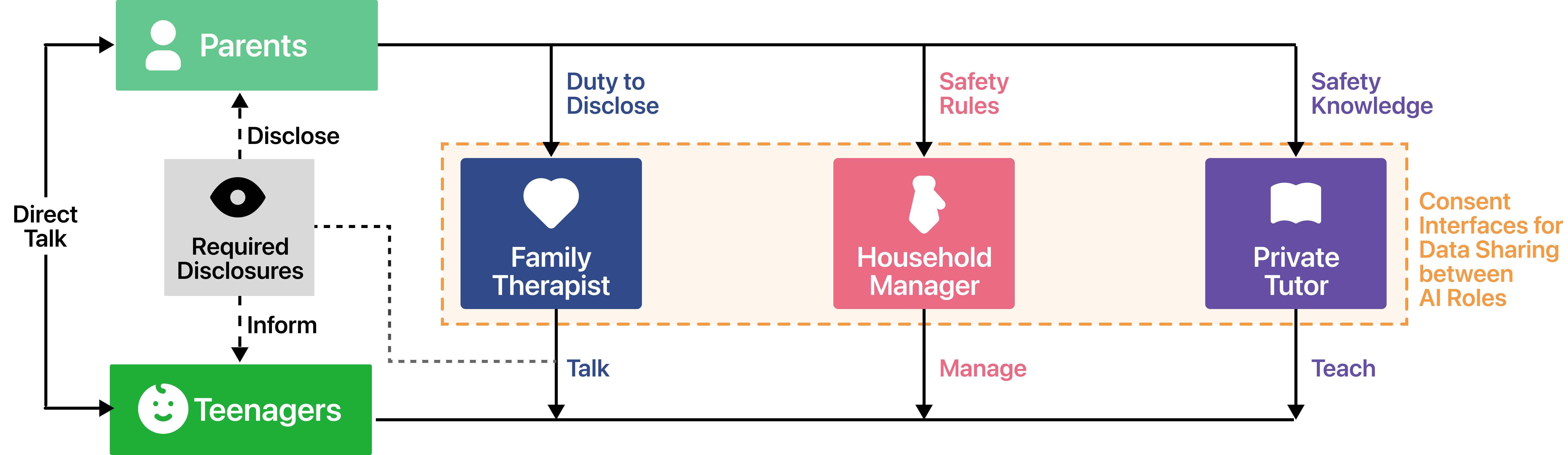}
  \caption{A multi-agent system for for families, where a single AI entity adopts multiple caregiver roles, where parents establish safety parameters and teens engage directly with each agent. Data sharing between AI agents requires family members consent.}
  \Description{This diagram draws a multi-agent system for families.}
  \label{fig:agents-family-safety}
\end{figure}

Based on our findings, we propose a multi-agent system for families that integrates three caregiving agents: household manager, private tutor, and family therapist. These agents support daily family routines while respecting family privacy boundaries and communication preferences. We first link these agent design elements to our study findings, then illustrate how each agent, as depicted in Figure~\ref{fig:agents-family-safety}, enhances rather than replaces parent-child communication about safety issues.

\textbf{Mapping Design Elements to Findings:} Families personified AI agents by envisioning three primary roles (Section~\ref{sec:roles}): a household manager assisting with daily tasks (e.g., chore tracking, managing emails), a private tutor offering on-demand academic help, and a family therapist providing emotional and relational guidance. Participants specifically mentioned needing support with tasks like managing email overload, homework assistance, and addressing sensitive topics that are difficult to discuss directly with parents.

Participants frequently discussed safety in relation to these agents (Section~\ref{sec:safety-findings}). They wanted AI agents to mitigate risks like scams, cyberbullying, and unwanted contact without feeling invasive. Some teens wanted to discuss sensitive issues with the AI therapist first before talking to their parents. Thus, our design integrates safety features into each agent instead of isolating safety as a standalone agent.

Families preferred clear separation of data across agents unless explicitly permitted. Parents highlighted that details shared by children with the AI therapist can remain private unless significant risks emerged. Meanwhile, children valued the option to keep their secrets with AI agents. These insights from Section~\ref{sec:concerns} informed our design of private information-sharing principles.

\textbf{Agent Interactions Design:} Our multi-agent system for families includes three AI agents (i.e., family therapist, household manager, and private tutor). Figure~\ref{fig:agents-family-safety} illustrates the protocol of the AI agents' interactions:

\begin{itemize}
    \item \textit{Direct Talk Pathway (solid bidirectional arrow):} The \textbf{direct talk} arrow between parents and teens emphasizes AI's supplementary role, preserving essential face-to-face discussions for critical safety issues.
    
    \item \textit{Family Therapist (blue):} The family therapist needs to handle safety situations meeting the \textbf{duty to disclose} criteria set by parents. In these cases, the therapist follows the protocol: informing and preparing teenagers about the disclosure, then sharing safety-related concerns with parents. The diagram illustrates this via dotted lines through \textbf{required disclosures}, emphasizing teenage autonomy alongside safety needs. Rather than directly reporting to parents, the therapist supports teenagers in leading sensitive discussions and providing guidance and assistance. Only if immediate safety concerns emerge does the therapist disclose information directly to parents, always keeping teenagers informed and involved through the \textit{talk} pathway.
    
    \item \textit{Household Manager (pink):} The household manager embeds \textbf{safety rules} within everyday activities through the \textit{manage} pathway, such as email management and household organization, ensuring safety protection without intrusive surveillance.

    \item \textit{Private Tutor (purple):} The private tutor includes \textbf{safety knowledge} education in academic support. Through the \textit{teach} pathway, this role offers educational assistance and safety knowledge. Its goal is to serve as an ever-present teacher, helping teens develop practical life skills, especially when parents are occupied.
    
    \item \textit{Consent Interfaces for Data Sharing (orange dashed line)}: The \textbf{consent interfaces} ensure data collected by each agent remains private unless explicitly permitted for sharing, respecting families' preference for agent-specific privacy and preventing unintended exposure of sensitive information.
\end{itemize}

This framework aims to address key roles, tensions, and safety concerns identified in our study so that it can protect family safety while respecting teens' autonomy.

\textbf{Illustrative Scenario for Framework Operation:} To illustrate how these three AI caregiving roles interact, we present a scenario highlighting the distinct engagement pathways (\emph{Talk, Manage, Teach}), the duty to disclose process, and role-specific data sharing protocols. This scenario addresses the safety issue of a teenager named T overwhelmed by upcoming exams and hurtful rumors spreading about them in an online class forum.
 
\begin{itemize}

\item \emph{Household Manager (Manage):} The household manager organizes emails from online class forums, helps with exam preparation, and sends reminders. Getting messages about hurtful rumors on T's laptop, the manager monitors suspicious emails or forum posts based on preset \textbf{safety rules}. Upon identifying potentially harmful content, the manager prompts T to discuss these issues with the therapist, encouraging T to seek support. Once T identifies content as harmful, the manager blocks similar content based on T’s preferences.

\item \emph{Family Therapist (Talk):} When T shares feelings of harassment or depression, the therapist invites T to explore coping strategies. If T's situation meets the \textbf{duty to disclose} threshold, such as mentioning self-harm, the therapist clearly explains the need to involve parents. The therapist assists T in preparing for this conversation, encouraging \textbf{direct talk} between parent and T. After a preset agreed period, if T hasn't spoken to parents, the therapist discloses the concern while keeping T informed. The therapist also offers to facilitate the discussion or involve the tutor for specific guidance on handling online rumors and effective reporting strategies. If T agrees to involve the tutor, the therapist explicitly seeks T’s consent before sharing relevant discussion details.

\item \emph{Private Tutor (Teach):} If T or the parents request advice on managing online rumors and effective reporting methods, the private tutor initially provides general \textbf{safety knowledge}. If additional personalization is beneficial, the tutor explicitly asks T if it can access previous conversations T had with the therapist. The tutor waits for clear consent before retrieving any sensitive details from the therapist.

\end{itemize}

In this scenario, the family therapist, household manager, and private tutor each maintain distinct privacy boundaries while coordinating via \textbf{consent interfaces}. Processes like the \textbf{duty to disclose} reflect family-defined thresholds, ensuring that teens maintain autonomy until genuine risks emerge. These AI agents enhance rather than replace parent-child communication, encourage open dialogue, and naturally integrate safety into everyday routines.

\subsubsection{Privacy Management of AI Agent System}
\label{sec:cpm-extend}
Building upon our framework's foundation, we now elaborate on the design of privacy and consent mechanisms. Our approach addresses families' explicit need for clear boundaries between AI agents. As P05 (45-54, male) noted, \textit{``Maybe is the information that another one has the right to keep private, and so just like because therapist knows it doesn't mean necessarily that others have the same privilege of knowing it.''}

The insights from our findings led to our proposal of four privacy management principles: 1) memory segregation to keep information distinct across AI agents, 2) conversational consent to ensure user-informed agreement on data sharing between AI agents, 3) selective sharing to provide fine-grained control over data dissemination, and 4) progressive memory to adapt data retention to evolving family relationships. These principles reflect research on CPM theory~\cite{petronio_communication_2010} that guided the negotiated disclosure paradigm and research on collaborative family decision-making around sensitive data sharing~\cite{dumaru2024s}.

At the foundation level, \textbf{memory segregation} ensures that each agent maintains distinct memory boundaries, just as different professionals maintain distinct relationships with family members. This principle directly responds to family concerns about maintaining clear information boundaries. For instance, it ensures conversations with an AI therapist remain as confidential as interactions with a human therapist, preventing information from being shared with other AI agents without explicit consent.

Building upon the foundation, \textbf{conversational consent} engages family members in natural conversations to seek explicit permission for sharing safety-relevant information. For example, the therapist would initiate a discussion with a teenager, \textit{``I notice this is an important safety concern. Could I share some of these insights with the household manager to help keep you safe?''} This approach ensures family members, especially teenagers, maintain agency over their information while understanding the safety implications.

Once consent is established, \textbf{selective sharing} guides agents to share data by extracting and communicating only specific, relevant insights necessary for safety and support purposes. This approach minimizes exposure by ensuring that sensitive details remain confined to their original context while enabling necessary coordination among AI agents.

To manage data sustainably, \textbf{progressive memory} governs how long the AI agent system retains various types of user data. Instead of keeping all data indefinitely, the system progressively dismisses information while only preserving safety-related data, thus addressing family desires for long-term privacy protection alongside support for household safety.

Together, these four principles collectively create a privacy management framework that is intuitive and controllable for families. Our framework supports families in naturally managing their own information shared with the multi-agent system, thereby maximizing the benefits of AI agents in enhancing integrated family safety and privacy.

\subsection{Ethical Considerations and Risks of AI Agent Integration}

While our multi-agent system design provides integrated and relational multi-agent safety support for families, it may introduce notable risks~\cite{timmons2023call,kretzschmar2019can,zhai2024effects,wang2024challenges}. One major concern is the illusion of emotional safety that we mentioned in the discussion: teens may perceive an AI therapist as empathetic, although agents cannot truly understand or respond to complex emotional needs~\cite{timmons2023call}. This false sense of support may delay help-seeking or lead to misplaced trust.

Context-agnostic models~\cite{kretzschmar2019can} may also misread family norms, offering advice that misaligns with family values or violates privacy boundaries. Families may also become overly reliant on AI agents~\cite{zhai2024effects}, diminishing the role of direct human judgment and emotional intuition that is often essential in safety-critical situations. Finally, delegating safety responsibilities to AI could shift accountability away from parents~\cite{wang2024challenges}, weakening family agency. These risks highlight the need for ongoing evaluation and technology corrections to protect against unintended consequences.

\section{Limitations}
Our study has four main limitations. Firstly, our sample was relatively homogeneous, consisting mainly of middle to upper-middle-class families from urban or suburban areas in the U.S. This limits the ability to generalize our findings to families from other socioeconomic backgrounds or cultural contexts. Future research may include families from diverse backgrounds to explore varying conceptualizations of the AI agents and their corresponding safety needs.

Secondly, our findings are based on families' imagined interactions with AI agents rather than actual usage. While this speculative approach allowed us to explore future possibilities, it may not fully capture the challenges that would emerge in real-world implementations.

Thirdly, we focused on parent-child dyads with teenagers between 13 and 15 years old, excluding younger children and older adolescents. The dynamics and safety needs may differ significantly for other age groups. Additionally, including only one child per family may not capture the full complexity of household with multiple children.

Finally, our research was conducted during a period of rapid GenAI advancement and shifting public perceptions of AI agent capabilities. Families' attitudes and expectations may evolve as they gain more experience with AI technologies. Future studies may track how these perceptions change over time with increased exposure to AI.

\section{Conclusion}
As families face growing digital and physical safety threats, our study highlights how generative AI agents can support household safety by embedding protective functions into family caregiving roles (e.g., household manager, private tutor, and family therapist). Rather than adopting a standalone safety AI agent, families preferred AI agent systems that support daily tasks, respect family privacy boundaries, and foster open parent-child safety communication. We propose a multi-agent system for families that balances safety and privacy needs by designing privacy management principles for the system, including memory segregation, conversational consent, selective sharing, and progressive memory. Future research should explore real-world deployments and adaptive safety mechanisms that align with diverse family dynamics and changing expectations.

\begin{acks}
The authors are grateful to the families whose participation made this research possible and to the anonymous reviewers for their valuable feedback. This research is in part supported by the National Science Foundation CNS-2341187, CNS-2426397, CNS-2442221, a Google PSS Faculty Award, and a Meta Research Award.
\end{acks}

\bibliographystyle{ACM-Reference-Format}
\bibliography{family-ai}

\appendix
\section{Appendix: Phase 1 Study Procedure Protocol}
\label{app:interview-script}

\subsection{Parent Session}
{\small
\subsubsection{Introduction and Warm-up (3 minutes)}

Thanks for participating in our study! Today, we'll be exploring ChatGPT together. Before we dive into the hands-on session, I'd like to briefly explain what ChatGPT is. ChatGPT is a type of generative AI tool. Have you heard about generative AI recently? 

Generative AI is shortened to GenAI and it refers to systems that can create new content, such as text messages, voice messages, or images, based on prompts you provide. 

It is important to note that although ChatGPT's responses may seem logical, it actually works by probabilistically popping out each word one by one based on what it learns from very large datasets that comprise billions of human-generated data records. So the ChatGPT outputs can be inaccurate, biased, or even inappropriate. 

If you encounter anything during the hands-on session that makes you uncomfortable, you can let me know. And you can pause and stop at any time in our study. Before we move on to the hands-on activities, do you have any questions about ChatGPT or generative AI?

\subsubsection{ChatGPT Hands-on Activities (10 minutes)}

Now, you are going to explore ChatGPT in text, image, and voice modes freely. And we will do this one by one. (The researcher opens a blank chat.)

\begin{enumerate}
    \item Text Mode: ``The first one is text mode. You can type any question or request you like. For example, you can ask for parenting tips, but the choice is yours.''
    \item Image Mode: ``In addition to text-based interaction, you can also type a prompt that asks ChatGPT to create or edit an image. Feel free to request to generate and edit any picture that interests you.''
    \item Voice Mode: ``Now, I'm going to show you the voice conversation mode. (The researcher shows the four voice options provided by ChatGPT.) You can click and hear these voices and choose the one you prefer. When you press the voice button in this bottom right corner, you can speak to ChatGPT. You may choose any topic or follow up on the previous interaction you already had with ChatGPT.''
    \item Wrap-up: After two or three turns in voice mode, the researcher concludes, ``Let's pause here so we have time to discuss your impressions.'' 
\end{enumerate}

\subsubsection{Impression Interview Questions (7 minutes)}

\paragraph{Immediate Reactions}
\begin{enumerate}[label*=\arabic*.]
    \item Did ChatGPT give you the kind of responses you were hoping for?
    \item Why or why not?
    \item Do you have any safety concerns about using ChatGPT?
\end{enumerate} 

\paragraph{Prior GenAI Experience}
\begin{enumerate}[label*=\arabic*.]
    \item Have you or your family used any AI tools that can generate answers, suggestions, or other content, such as ChatGPT, email auto-complete, or smart home devices like Alexa or Siri?
    \begin{enumerate}[label*=\alph*.]
        \item If not, that is totally fine.
        \item If yes, what kinds of things do you usually use them for? Do you have any safety concerns about them?
    \end{enumerate}
\end{enumerate}
}
\subsection{Child Session}

{\small

\subsubsection{Introduction and Warm-up (3 minutes)}

Thanks for helping with our research! Today, we will try out ChatGPT together. Before we start, let me explain very quickly what ChatGPT is. ChatGPT is a kind of computer tool called generative AI. Have you heard about generative AI recently? 

Generative AI means that it can make new things like words and pictures or talk like a chat bot. It makes new things after you give it a prompt. A prompt is what you type or say to the computer tool to get it started. 

ChatGPT decides what to show you by guessing word-by-word from a huge amount of information that people have created on the Internet. Because it is guessing, it can sometimes be wrong, confusing, or even output things that you do not feel comfortable. 

If you feel uncomfortable with ChatGPT's outputs, you can tell me to stop. You can also tell me at any time during the study that you want to stop. Before we jump in, do you have any questions about what I just said? Like, about ChatGPT and generative AI?

\subsubsection{ChatGPT Hands-on Activities (10 minutes)}

Now, you will interact with ChatGPT in three ways: typing, creating images, and talking. And we will do this one by one. (The researcher opens a blank chat.)

\begin{enumerate}
    \item Text Mode: ``First is typing. You can ask anything you are curious about. Some kids ask it to tell a joke, but it is totally up to you.''
    \item Image Mode: ``Next is creating images. You can type a prompt that tells it to draw something. Anything that you want ChatGPT to draw?''
    \item Voice Mode: ``Now the last one is talking. Here are four voice choices. (The researcher shows the four voice options provided by ChatGPT.) You can pick the one you like. Then when you press the voice button in this bottom right corner, you can talk to ChatGPT about any topic or continue what you were doing with ChatGPT.''
    \item Wrap-up: After two or three turns in voice mode, the researcher concludes, ``OK! Let's pause here so we can chat about what you thought.''
\end{enumerate}

\subsubsection{Impression Interview Questions (7 minutes)}

\paragraph{Immediate Reactions}
\begin{enumerate}[label*=\arabic*.]
    \item Did ChatGPT give you the kind of answers or pictures you wanted?
    \item Why or why not?
    \item Did anything about ChatGPT make you feel unsure or worried?
\end{enumerate}

\paragraph{Prior GenAI Experience}
\begin{enumerate}[label*=\arabic*.]
    \item Have you ever used something that can answer your questions or make things for you, like words or pictures? For example, like writing helpers or smart home devices like Alexa or Siri?
    \begin{enumerate}[label*=\alph*.]
      \item If not, that is totally fine.
      \item If yes, what do you like using them for? Did any of these tools ever make you feel unsure or worried?
    \end{enumerate}
\end{enumerate}
}

\section{Appendix: Phase 2 Codebook}

\begingroup
\footnotesize

\begin{longtable}[t]{@{}
  >{\RaggedRight\arraybackslash}p{0.12\textwidth}  
  >{\RaggedRight\arraybackslash}p{0.17\textwidth}  
  >{\RaggedRight\arraybackslash}p{0.65\textwidth}  
@{}}


\toprule
\textbf{Themes} & \textbf{Codes} & \textbf{Example Quotes} \\
\midrule
\endfirsthead

\toprule
\textbf{Themes} & \textbf{Codes} & \textbf{Quotations} \\
\midrule
\endhead

\bottomrule
\multicolumn{3}{r}{\textit{Continued on next page...}}\\
\endfoot

\bottomrule
\endlastfoot

\multirow[t]{3}{0.12\textwidth}{Envisioned Roles of AI Agents}

& Household Manager &
\begin{minipage}[t]{\linewidth}
\begin{itemize}[leftmargin=8pt, itemsep=2pt]
  \item[-] C05: ``AI should help with processing emails... you might be able to do some things such as like hanging out with us more.''
  \item[-] C10: \enquote{I cannot wake up for the life of me... AI can push me like \enquote{[C10's name], it's time to get up.}}
\end{itemize}
\end{minipage}
\\
\cmidrule(lr){2-3}
\noalign{\vskip 1pt}

& Private Tutor &
\begin{minipage}[t]{\linewidth}
\begin{itemize}[leftmargin=8pt, itemsep=2pt]
  \item[-] C07: ``[AI] can explain things in more than one way.''
  \item[-] C08: ``School's not really working out for me... my teacher doesn't really explain it that well.''
\end{itemize}
\end{minipage}
\\
\cmidrule(lr){2-3}
\noalign{\vskip 1pt}

& Family Therapist &
\begin{minipage}[t]{\linewidth}
\begin{itemize}[leftmargin=8pt, itemsep=2pt]
  \item[-] C05: ``[AI could help with] how did I deal with [sibling fights] before?''
  \item[-] C04: ``AI would be great... I don't specifically talk to my parents about messages [with friends] sometimes. Because it's sometimes difficult to talk to my parents about [it].''
  \item[-] P13: ``A Gen Z and Gen X translator... like someone pulling out old letters from the attic...''
\end{itemize}
\end{minipage}
\\

\midrule

\multirow[t]{3}{0.12\textwidth}{Safety Feature Integration}

& Embedded Safety Risk Detection &
\begin{minipage}[t]{\linewidth}
\begin{itemize}[leftmargin=8pt, itemsep=2pt]
  \item[-] C02: ``Forget to lock your doors... or forget to turn off something like the stove.''
  \item[-] C04: ``Like my friend coughs up blood... we can ask AI if it's OK that I'm coughing up blood?''
  \item[-] P11: ``Parents should get an alert if the child is buying alcohol [online].''
\end{itemize}
\end{minipage}
\\
\cmidrule(lr){2-3}
\noalign{\vskip 1pt}

& Critical Thinking and Safety Skills &
\begin{minipage}[t]{\linewidth}
\begin{itemize}[leftmargin=8pt, itemsep=2pt]
  \item[-] P06: \enquote{It's getting harder and harder to tell what's real and what's not out in the news world. So that can be a useful teacher on, \enquote{Hey, I heard this story...}}
  \item[-] P06: ``Well, you're gonna start driving before too long... as you get prepared for that and you know, tips on safe driving.''
\end{itemize}
\end{minipage}
\\
\cmidrule(lr){2-3}
\noalign{\vskip 1pt}

& Mental Health and Safety Conversations &
\begin{minipage}[t]{\linewidth}
\begin{itemize}[leftmargin=8pt, itemsep=2pt]
  \item[-] C05: \enquote{[AI] could ask... \enquote{you sure you wanna send this?}}
  \item[-] P02: \enquote{[AI is] that silent person... and basically says, \enquote{This is not OK.}}
  \item[-] P03: ``There's just a lot about my boy's inner world that I won't know.''
\end{itemize}
\end{minipage}
\\

\midrule

\multirow[t]{3}{0.12\textwidth}{Adoption Concerns and Needs}

& Mental Model \par Differences &
\begin{minipage}[t]{\linewidth}
\begin{itemize}[leftmargin=8pt, itemsep=2pt]
  \item[-] C01: ``AI won't judge me. Whatever happened.''
  \item[-] P13: ``It's creepy how advanced AI is... it's invasive.''
\end{itemize}
\end{minipage}
\\
\cmidrule(lr){2-3}
\noalign{\vskip 1pt}

& Privacy Boundaries &
\begin{minipage}[t]{\linewidth}
\begin{itemize}[leftmargin=8pt, itemsep=2pt]
  \item[-] P05: ``Just because therapist knows it doesn't mean others have the same privilege of knowing it.''
\end{itemize}
\end{minipage}
\\
\cmidrule(lr){2-3}
\noalign{\vskip 1pt}

& Direct \par Communication &
\begin{minipage}[t]{\linewidth}
\begin{itemize}[leftmargin=8pt, itemsep=2pt]
  \item[-] P05: ``In the end it should be that [C05's name] and I are communicating [about safety] not through the AI.''
\end{itemize}
\end{minipage}
\\

\end{longtable}
\endgroup

\end{document}